\documentclass[iop,letter]{emulateapj}
\usepackage{psfig}
\usepackage{apjfonts}
\usepackage{ulem}
\usepackage[dvips]{color}
\usepackage{graphicx}
\usepackage[USenglish]{babel}
\usepackage{mathrsfs}
\def\stan{\sigma_{\rm tan}}
\def\srad{\sigma_{\rm rad}}
\begin{document}

\title{The Hubble Space Telescope UV Legacy Survey of galactic
  globular clusters. VI. The internal kinematics of the multiple
  stellar populations in NGC~2808$^{\ast}$}\footnotetext[$^{\ast}$]
      {Based on observations with the NASA/ESA \textit{Hubble
          Space Telescope}, obtained at the Space Telescope Science
        Institute, which is operated by AURA, Inc., under NASA
        contract NAS 5-26555.}

\author{
A.\ Bellini\altaffilmark{1},
E.\ Vesperini\altaffilmark{2},
G.\ Piotto\altaffilmark{3,4},
A.\ P.\ Milone\altaffilmark{5},
J.\ Hong\altaffilmark{2},
J. Anderson\altaffilmark{1},
R.\ P.\ van der Marel\altaffilmark{1},
L.\ R.\ Bedin\altaffilmark{4},
S.\ Cassisi\altaffilmark{6,7},
F.\ D'Antona\altaffilmark{8},
A.\ F.\ Marino\altaffilmark{5},
A.\ Renzini\altaffilmark{4}
}

\altaffiltext{1}{Space Telescope Science Institute, 3700 San Martin
  Dr., Baltimore, MD 21218, USA}

\altaffiltext{2}{Department of Astronomy, Indiana University,
  Bloomington, IN 47405, USA}

\altaffiltext{3}{Dipartimento di Fisica e Astronomia ``Galileo
  Galilei'', Universit\`a di Padova, v.co dell'Osservatorio 3,
  I-35122, Padova, Italy}

\altaffiltext{4}{Istituto Nazionale di Astrofisica, Osservatorio
  Astronomico di Padova, v.co dell'Osservatorio 5, I-35122, Padova,
  Italy}

\altaffiltext{5}{Research School of Astronomy \& Astrophysics,
  Australian National University, Mt Stromlo Observatory, via Cotter
  Rd, Weston, ACT 2611, Australia}

\altaffiltext{6}{Istituto Nazionale di Astrofisica, Osservatorio
  Astronomico di Teramo, Via Mentore Maggini s.n.c., I-64100, Teramo,
  Italy}

\altaffiltext{7}{Instituto de Astrof\`isica de Canarias, E-38200 La
  Laguna, Tenerife, Canary Islands, Spain}

\altaffiltext{8}{Istituto Nazionale di Astrofisica, Osservatorio
  Astronomico di Roma, Via Frascati 33, I-00040, Monteporzio Catone,
  Roma, Italy}

\topmargin 1.0cm

\begin{abstract} 
  Numerous observational studies have revealed the ubiquitous presence
  of multiple stellar populations in globular clusters and cast many
  hard challenges for the study of the formation and dynamical history
  of these stellar systems. In this Letter we present the results of a
  study of the kinematic properties of multiple populations in
  NGC~2808 based on high-precision \textit{Hubble Space Telescope}
  proper-motion measurements. In a recent study, Milone et al. have
  identified five distinct populations (A, B, C, D, and E) in
  NGC~2808.  Populations D and E coincide with the helium-enhanced
  populations in the middle and the blue main sequences (mMS and bMS)
  previously discovered by Piotto et al.; populations A, B, and C
  correspond to the redder main sequence (rMS) that in the Piotto et
  al. was associated with the primordial stellar population.  Our
  analysis shows that, in the outermost regions probed (between about
  1.5 and 2 times the cluster half-light radius), the velocity
  distribution of populations D and E is radially anisotropic (the
  deviation from an isotropic distribution is significant at the
  $\sim$3.5-$\sigma$ level).  Stars of populations D and E have a
  smaller tangential velocity dispersion than those of populations A,
  B, and C, while no significant differences are found in the
  radial-velocity dispersion.  We present the results of a numerical
  simulation showing that the observed differences between the
  kinematics of these stellar populations are consistent with the
  expected kinematic fingerprint of the diffusion towards the cluster
  outer regions of stellar populations initially more centrally
  concentrated.
\end{abstract}


\maketitle

\section{Introduction}
\label{sec:intro}

Photometric and spectroscopic studies over the last 15 years have
revolutionized our understanding of globular clusters (GCs). Once
thought to be ``simple stellar populations'' with a single age and
composition, essentially all GCs host multiple stellar populations
(MSPs), as revealed by their multiple photometric sequences.  The
exquisite precision and stability of the \textit{Hubble Space
  Telescope} (\textit{HST}) has made much of this possible (see
\citealt{2015MNRAS.447..927M, 2015ApJcc, 2015arXiv150407876N,
  2015AJ....149...91P}, and references therein).  MSPs can be traced
to different compositions in terms of light elements (such as Na, O,
Al, Mg observed spectroscopically; see \citealt{2012A&ARv..20...50G}
and references therein) and He \citealt{2011A&A...531A..35P};
\citealt{2012ApJ...748...62V}; \citealt{2014MNRAS.437.1609M}). These
observational findings have cast a number of challenges for the study
of the formation and evolution of GCs.

In order to make progress toward a complete picture of the formation
and evolutionary history of GCs, as well as to constrain the possible
paths for the theoretical study of these complex challenges and
questions, a major effort to combine observational data is essential.
Spectroscopic and photometric studies have begun to shed light on the
chemical properties, number of distinct sequences and range of ages of
different stellar populations.

Structural and kinematic properties of different stellar populations
are two additional key pieces of the puzzle, as they contain essential
information to build a complete picture of MSP cluster formation and
dynamical evolution.  According to a number of different formation
scenarios (see, e.g., \citealt{dercole2008}; \citealt{decressin};
\citealt{bastian}), second-generation (2G) populations should form
more concentrated in the cluster inner regions\footnote[1]{Hereafter,
  we will use ``populations'' and ``generations'' as synonyms.}
Numerical simulations \citep{vesperinietal2013} have shown that
several clusters should still retain some memory of this initial
spatial segregation and a few observational studies have indeed found
clusters in which 2G stars are still concentrated in the cluster inner
regions \citep{2007ApJ...654..915S, 2009A&A...507.1393B, johnson,
  2012A&A...537A..77M, 2011A&A...525A.114L, 2013ApJ...765...32B,
  cordero}.

Little is however known to-date on the proper-motion (PM) based
kinematics of MSPs (with the only exceptions of $\omega$~Cen and
47~Tuc:\ \citealt{2009A&A...493..959B, 2010ApJ...710.1032A,
  2013ApJ...771L..15R}).  Richer et al. (2013) probed the outer
regions of 47 Tuc (at $\sim$1.9\,$r_{\rm h}$) and found that 2G stars
are characterized by a radially anisotropic velocity distribution
while the 1G population is isotropic.  As discussed in
Section~\ref{sec:theo}, this trend is consistent with the expected
kinematical implications of the formation models mentioned above; the
study presented in this Letter is aimed at exploring whether the
kinematics of MSPs in NGC~2808 further confirms this general
expectation and the trend identified in 47 Tuc.

NGC~2808 is one of the most massive Galactic GCs and one of the most
massive Galactic GCs and one of the few known to host a super-He-rich
subpopulation (see, e.g., \citealt{2005ApJ...631..868D,
  2007ApJ...661L..53P}).  The initial picture of three distinct
stellar populations found by \citet{2007ApJ...661L..53P} has been
recently shown to be even more complex by \citet{2015ApJcc}, who
identified five distinct populations (named A, B, C, D and E) along
the red-giant branch (RGB) and the main sequence (MS) of NGC~2808. We
point out here that populations D and E correspond to the 2G
populations identified by \citet[their mMS and
  bMS]{2007ApJ...661L..53P}. Populations A, B, and C correspond to the
rMS of the \citet{2007ApJ...661L..53P} study (that they assumed to be
the first generation, 1G).  New WFC3/UVIS UV data (GO-12605, PI:
Piotto) allowed us to further split the rMS into the three populations
A, B and C. Available spectroscopic data discussed in
\citet{2015ApJcc} suggest that population B has Na and O abundances
usually associated with 1G stars; population C includes stars with
slightly enhanced Na and slightly depleted O abundances, and is
photometrically distinct from population B along the RGB (see
\citealt[their Fig.~8]{2015ApJcc}).  No spectroscopic data are
available to further characterize population A.

In this Letter we shed further light on the properties of this complex
cluster and present an analysis of the internal PMs of its five
stellar populations.  The outline of this Letter is as follows:\ in
Section~\ref{sec:dataset} we describe the data sets and reduction; in
Section~\ref{sec:obs} we present our results on the kinematic
properties of the five populations identified in NGC~2808. In
Section~\ref{sec:theo} we discuss the interpretation of the
observational findings. We summarize our conclusions in
Section~\ref{sec:conclu}.

\begin{figure}[t!]
\centering\includegraphics[width=7.9cm]{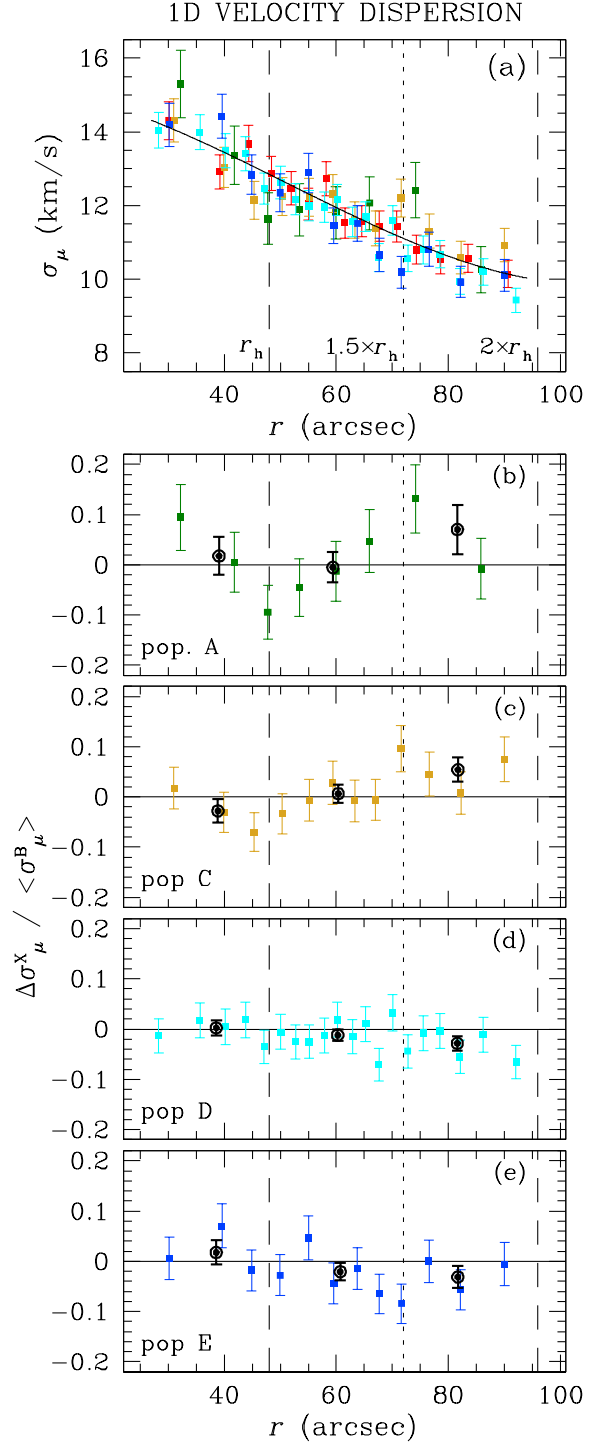}\\
\caption{(a) 1D $\sigma_\mu$ vs.\ $r$ for the 5 distinct
  populations. We least-squares fitted a 3rd-order polynomial to the
  population B profile ($\langle \sigma_\mu^{\rm B} \rangle$) (black
  curve). The half\_light radius, $1.5\times r_{\rm h}$, and $2\times
  r_{\rm h}$ (dashed and dotted lines) are also indicated. Panels (b)
  to (e) show the normalized difference between $\sigma_\mu$ of
  populations A, C, D, E, respectively, with respect to that of the
  reference population B. Colored points refer to bins containing the
  same number of stars per population. Black points refer to 3 fixed
  radial intervals:\ (1) $r\le r_{\rm h}$, (2) $r_{\rm h}<r\le
  1.5\times r_{\rm h}$, and (3) $1.5\times r_{\rm h}<r\le 2\times
  r_{\rm h}$.  See the text for details.}
\label{f1}
\end{figure}

\begin{figure}[t!]
\centering\includegraphics[width=7.9cm]{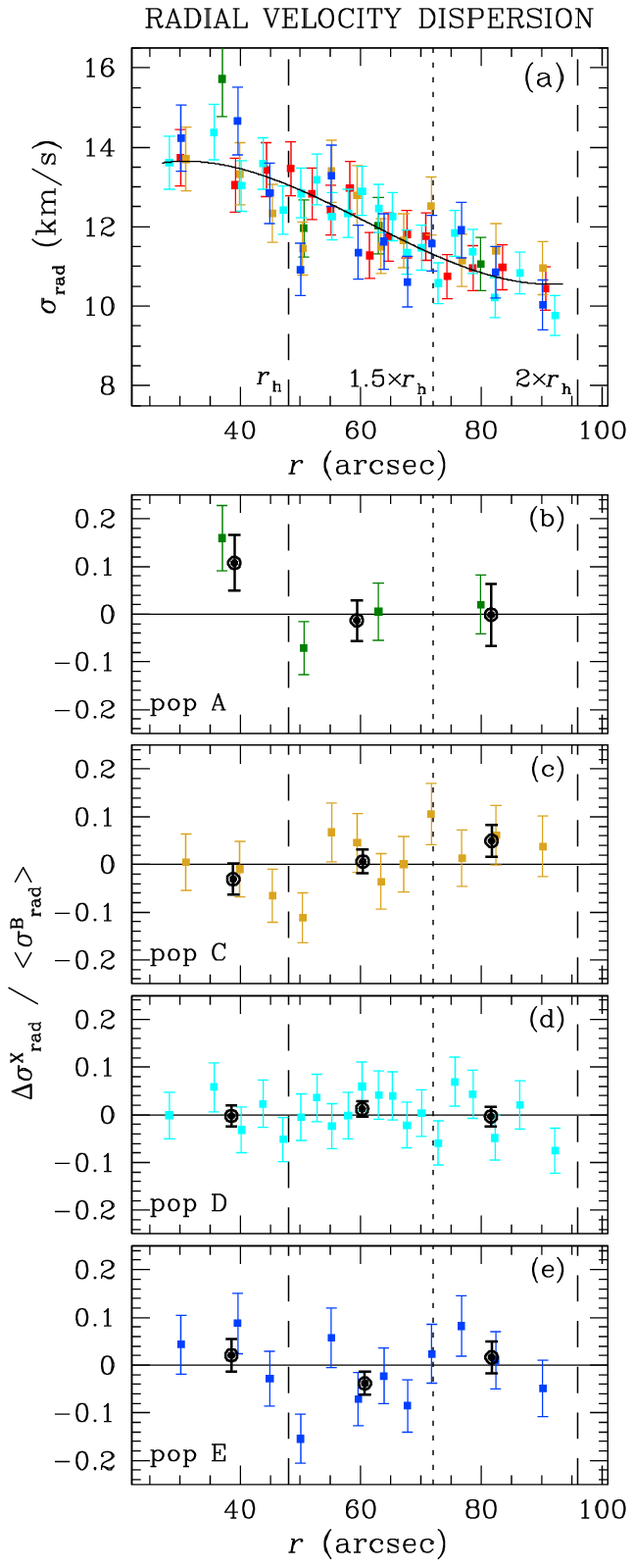}\\
\caption{Similar to Fig.~\ref{f1} but for the radial component of the
  motion. See the text for details.}
\label{f2}
\end{figure}

\begin{figure}[t!]
\centering \includegraphics[width=7.9cm]{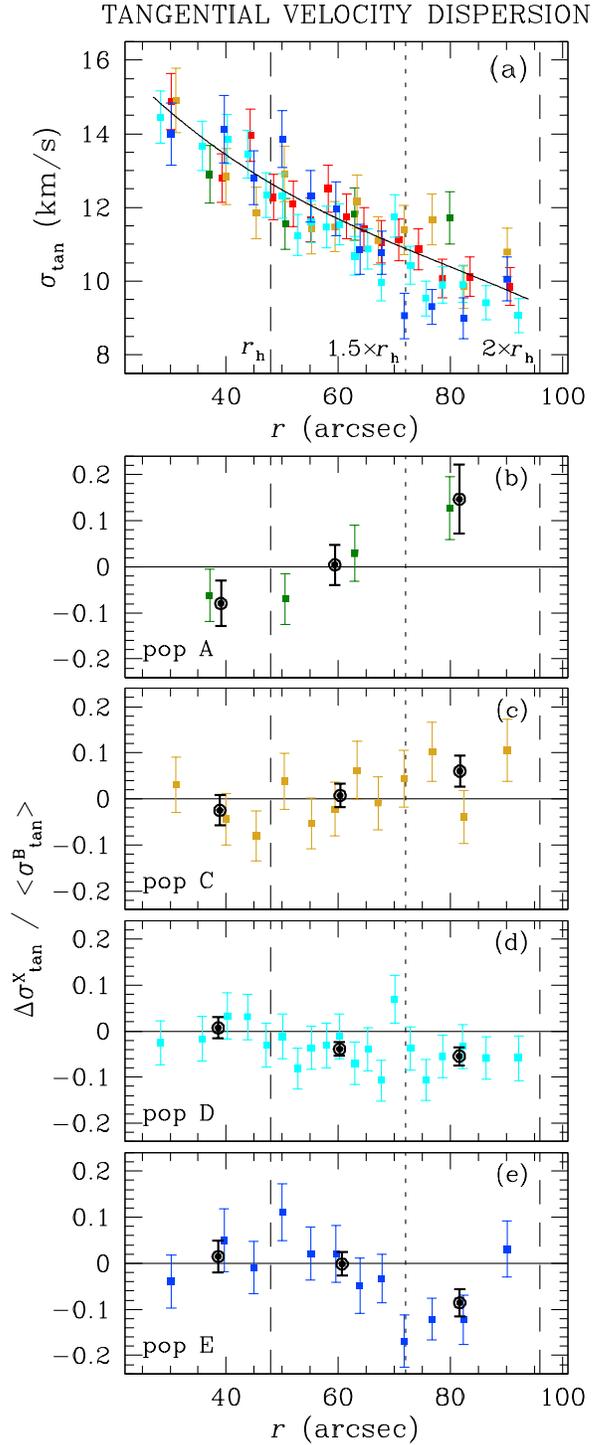}\\ 
\caption{Similar to Fig.~\ref{f1} but for the tangential component of
  the motion. See the text for details.}
\label{f3}
\end{figure}

\section{Data sets and Reduction}
\label{sec:dataset}
We have compiled high-precision PM catalogs for the central regions of
22 GCs, mostly based on the archival \textit{HST} proposal AR-12845
(P.I.: Bellini, see \citealt{2014ApJ...797..115B}).  PMs are obtained
following the \textit{central overlap} method
\citep{1971PMcCO..16..267E}, via a careful data-rejection
process. Extensive simulations have demonstrated the reliability of
the estimated PMs and their errors. The observations used to compute
PMs in NGC~2808 are listed in Table~11 of
\citet{2014ApJ...797..115B}. We additionally included WFC3/UVIS
observations of GO-12605 ($6\times 650\,$s in F336W and $6\times
97\,$s in F438W), taken in 2013, to further extend the available time
baseline and increase the PM accuracy. The catalog contains over
$86\,000$ stars down to $\sim 5$ mag below the turn off
(TO). Well-measured stars have a typical PM error of less than $\sim
30\, \mu$as ${\rm yr}^{-1}$ ($\sim$1.3 km$\,$s$^{-1}$, at a distance
of 9.6 kpc, \citealt{h96}).

The photometric catalog of \citet{2015ApJcc} combines the GO-10775
F606W and F814W ACS/WFC photometry presented in
\citet{2008AJ....135.2055A} with the WFC3/UVIS F275W, F336W and F438W
exposures of GO-12605. All observations were corrected for CTE effects
\citep{an10} and reduced using the software family \texttt{img2xym}
\citep{an06}, employing spatially-varying, time-dependent empirical
PSFs \citep{2013ApJ...769L..32B}. Stellar positions were corrected for
geometric distortion following the recipes in \citet{bb09} and
\citet{b11}. Photometry is calibrated as in
\citet{2005MNRAS.357.1038B}.

We refer to \citet{2014ApJ...797..115B} and \citet{2015ApJcc} for a
detailed description of the data reduction.

The cross-identification of common stars between the PM and the
photometric catalog left us with near $38\,000$ objects down to about
3 mag below the TO. In order to analyze in detail the internal
kinematics of NGC~2808 stars, particular care must be taken in
removing any PM measurements affected by systematic effects. To this
aim, we closely followed the recipes described in detail in
Section~7.5 of \citet{2014ApJ...797..115B}. Our final sample consists
of about $27\,000$ stars within $115^{\prime\prime}$ (the outermost
corner of the field of view) with high-precision, high-quality
measurements.

\section{The internal kinematics of the five subpopulations}
\label{sec:obs}

To avoid edge effects, we consider hereafter only stars within
$2\times r_{\rm h}= 96^{\prime\prime}$ (Harris 1996, 2010 edition).
Due to small-number statistics, the kinematics of the MSPs on the RGB
cannot be studied with sufficient detail.  On the other hand, we have
plenty of MS stars at our disposal, in particular:\ 598 population-A
stars, 3071 B, 1875 C, 4485 D, and 1814 E. We will use the same
nomenclature and color-coding of \citet{2015ApJcc}.

The 1D average transverse velocity-dispersion profile $\sigma_\mu$ of
each population, obtained by combining together
$\mu_{\alpha}\cos\delta$ and $\mu_\delta$, is shown in panel (a) of
Fig.~\ref{f1}. We binned each population in such a way as to have the
same number of measurements ($\gtrsim 150$) in each bin. We marked the
position of $r_{\rm h}$ (inner dashed line), $1.5\times r_{\rm h}$
(dotted line), and $2\times r_{\rm h}$ (outer dashed line).  All
errorbars refer to 1-$\sigma$ errors.

The five profiles largely follow the same trend. To better examine
small velocity-dispersion differences among them we proceeded as
follows. We chose population B (He-normal, 1G stars) as a reference,
and we least-squares fitted a 3rd-order polynomial to its $\sigma_\mu$
profile (red points in panel a) to obtain the average trend $\langle
\sigma_\mu^{\rm B}\rangle$.  Then, we computed the differences $\Delta
\sigma_\mu^{\rm X}$ between the $\sigma_\mu$ of the other 4
populations (${\rm X}={\rm A},{\rm C},{\rm D}, {\rm E}$) and the
average trend of population B at the same radial distance.  These
quantities are then normalized to the average trend of population B
itself. These normalized differences are shown in panels (b) to (e),
as a function of $r$, for populations A, C, D and E, respectively.

In addition, we binned each population into 3 radial intervals:\ (1)
$r\le r_{\rm h}$, (2) $r_{\rm h}<r\le 1.5\times r_{\rm h}$, and (3)
$1.5\times r_{\rm h}<r\le 2\times r_{\rm h}$. The respective values of
$\Delta \sigma_\mu^{\rm X}/\langle \sigma_\mu^{\rm B}\rangle$ for
these radial bins are shown as black points with
errorbars. Populations A and C (panels b and c) exhibit a slightly
increasing $\Delta \sigma_\mu^{\rm X}/\langle \sigma_\mu^{\rm
  B}\rangle$ values the larger the radial distance, while populations
D and E (panels d and e) seem to have an opposite behavior. These
differences are marginally significant (at the 2-$\sigma$ level).

We repeated the same analysis shown in Fig.~\ref{f1} independently for
the radial and tangential components of motion.\footnote[2]{We
  decreased the number of bins while keeping the same number of
  measurements within (at least 150) for populations A, D and E.}
Results are shown in Fig.~\ref{f2} and \ref{f3} for the radial
($\sigma_{\rm rad}$) and tangential ($\sigma_{\rm tan}$)
velocity-dispersion profiles, respectively.

First, we note the overall trend of $\sigma_{\rm tan}$
(Fig.~\ref{f3}a) being smaller than $\sigma_{\rm rad}$
(Fig.~\ref{f2}a) for $r\gtrsim r_{\rm h}$, in agreement with what
found by \citet{2015ApJ...803...29W} for the RGB of this cluster.  The
\textit{radial} velocity dispersions of the 5 populations do not show
significant differences (with only possible hints, at less than
$\sim$2-$\sigma$ level, of a larger $\srad$ of population A in the
innermost region, a larger $\srad$ for population C in the outermost
bin), while for $r>1.5\times r_{\rm h}$ the \textit{tangential}
velocity dispersion of populations D and E (panels d and e of
Fig.~\ref{f3}) is significantly smaller ($-0.054\pm0.019$ for D, and
$-0.086\pm0.028$ for E, i.e., at the 2.8- and 3.1-$\sigma$ level,
respectively) than that of the reference population B.  No significant
difference is instead found between the kinematics of population B and
that of A and C, with only a hint (at less than $\sim$2-$\sigma$
level) of a larger tangential velocity for populations A and C in the
outermost regions (panels b and c of Fig~\ref{f3})

\begin{figure}[t!]
\centering
\includegraphics[width=\columnwidth]{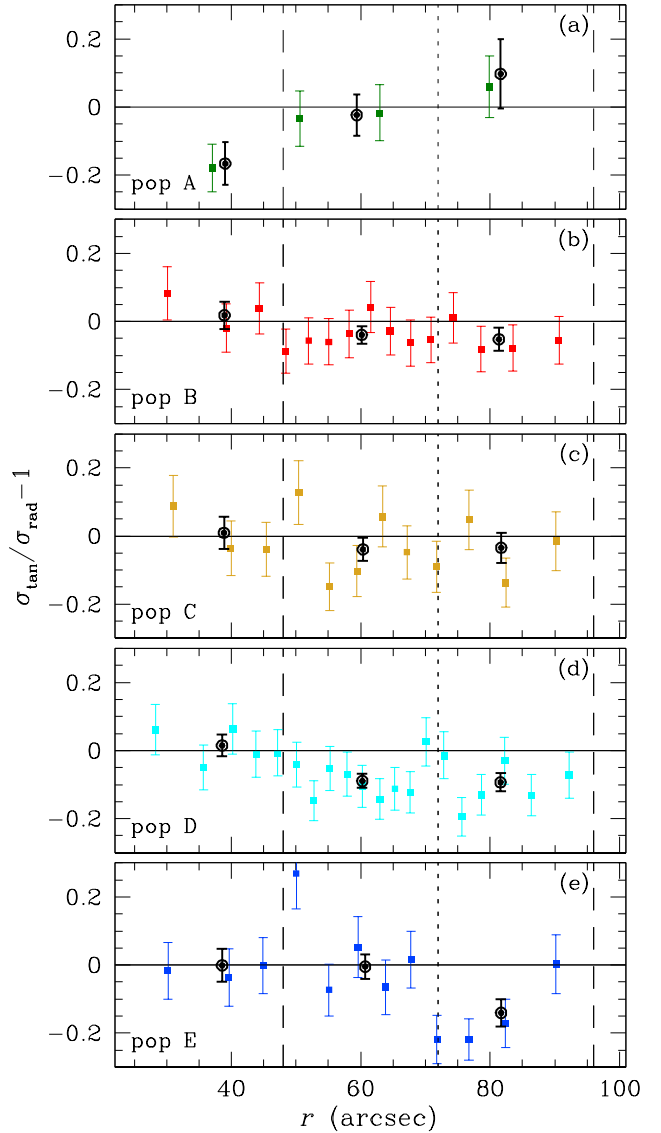}\\
\caption{Deviation from tangential-to-radial isotropy (horizontal
  line) for the 5 populations (color-coded as in
  Fig.~\ref{f1}). Vertical lines mark the locations of $r_{\rm h}$,
  $1.5\times r_{\rm h}$, and $2\times r_{\rm h}$.}
\label{f4}
\end{figure}

To further explore these findings, we computed the radial dependence
of the deviation from isotropy ($\sigma_{\rm tan}/\sigma_{\rm rad}-1$)
of each population.  Results are shown in Fig.~\ref{f4}.  The
horizontal line at 0 corresponds to an isotropic system.  As for
Fig.~\ref{f1}, larger black dots refer to data binned in the 3 radial
intervals, while colored points are used for data binned by keeping
the same number of stars per bin.

Populations B and C have an overall similar trend, and tend to very
slightly deviate from isotropy outside $r_{\rm h}$. In the outermost
regions probed (between $1.5r_{\rm h}$, and $2r_{\rm h}$) populations
D and E are radially anisotropic ($\sigma_{\rm tan}/\sigma_{\rm
  rad}-1$ equal to $-0.093\pm0.027$ for D and $-0.140\pm 0.040$ for
E:\ a deviation from an isotropic distribution at the $\sim$
3.4-$\sigma$ and 3.5-$\sigma$ level, respectively).  There is a hint
of population A having an opposite behavior with respect to the
others:\ the outer 2 points are fairly consistent with an isotropic
system, while the innermost point indicates a marginally-significant
(at the $\sim$2-$\sigma$ level) radial anisotropy for $r<r_{\rm
  h}$. As discussed in the Introduction, the nature of population A is
still unclear and spectroscopic data will be necessary to characterize
this population in the context of the MSPs of NGC~2808. Additional
kinematic data would also be necessary to shed further light on the
kinematic properties of this population and clarify the significance
of possible kinematic anomalies.

\section{Theoretical interpretation}
\label{sec:theo}

The two most statistically-significant results presented in the
previous sections are:\ i) populations D and E have a radially
anisotropic velocity distribution (Fig.~\ref{f4}) in the outermost
regions, and ii) this difference in anisotropy is due to the fact that
populations D and E have a smaller tangential velocity dispersion than
that of the other populations (Fig.~\ref{f3}). No strong differences
are instead found in the radial velocity dispersion of all the
populations (see Fig.~\ref{f2}).

One of the predictions of MSP cluster-formation models, in which
asymptotic-giant branch (AGB) stars are the source of gas out of which
2G stars form, is that the AGB ejecta would collect in the cluster
central regions and 2G stars would initially be more spatially
concentrated than the 1G population \citep{dercole2008}. Other
formation models, based on rapidly-rotating massive stars and/or
massive binaries (see e.g. \citealt{decressin, 2009A&A...507L...1D,
  bastian}) do not follow the evolution of the processed gas but
assume that massive stars are initially concentrated in the cluster
inner regions and so is the gas they release.  In any case, the
subsequent long-term dynamical evolution will gradually erase the
differences in the spatial distribution of 1G and 2G stars.

\begin{figure}[t!]
\centering
\includegraphics[width=\columnwidth]{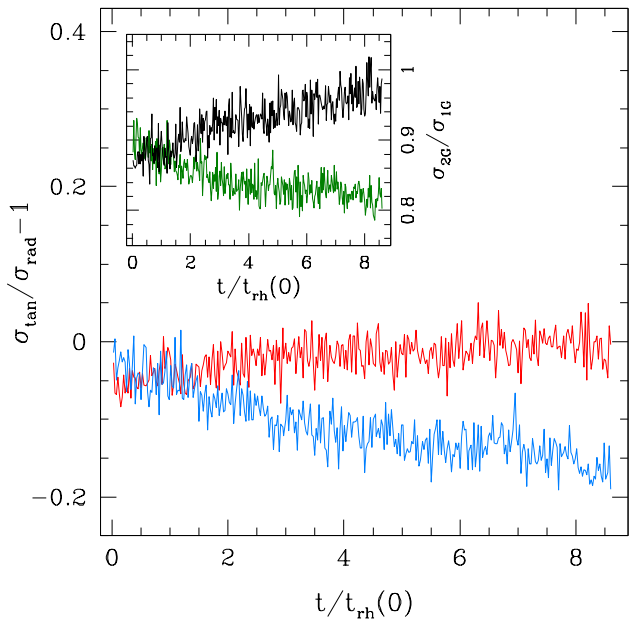}\\
\caption{Time evolution of $\stan/\srad-1$ measured for particles
  between 1.5 and 2 $r_{\rm h}$ for the 2G (blue line) and the 1G (red
  line) populations (where $r_{\rm h}$ is the projected half-mass
  radius). The green line in the inset shows the time evolution of the
  mean of the ratio of the 2G to the 1G tangential velocity
  dispersion, while the black line is for the mean of the ratio of the
  2G to the 1G radial velocity dispersion (both measured between 1.5
  and 2 $r_{\rm h}$). Time is normalized to the initial half-mass
  relaxation time (calculated using the projected half-mass radius).}
\label{f5}
\end{figure}

Here, we present the results of a N-body simulation carried out to
explore the evolution of the kinematical properties of different
stellar populations.  We started our simulation with 50$\,$000
equal-mass particles, with 1G and 2G populations having the same total
mass and both following a \citet{1966AJ.....71...64K} model density
profile with central dimensionless potential $W_0=7$. We assumed that
the 2G population is more centrally concentrated, with an initial
half-mass radius about 4.5 times smaller than that of the 1G
population. As for the kinematical properties, both 1G and 2G systems
are initially isotropic. The cluster is initially tidally limited and
is assumed to move on a circular orbit in the external potential of
the host galaxy (modeled as a point mass). We point out that we are
not ruling out the possibility that the formation process and the
subsequent virialization phase might result in 1G and 2G subsystems
with initially different anisotropy profiles, but here we focus solely
on the role of relaxation-driven long-term dynamical evolution in
establishing kinematical differences between different stellar
populations.  This simulation has been run with the GPU-accelerated
version of the NBODY6 code \citep{2003gnbs.book.....A,
  2012MNRAS.424..545N} on the Big Red II supercomputer at Indiana
University.

Observational data reveal that the strongest deviation from isotropy
occurs in the outermost regions for which data are
available. Figure~\ref{f5} shows the time evolution of $\stan/\srad-1$
measured for particles between 1.5 and 2 $r_{\rm h}$ (where $r_{\rm
  h}$ is the projected half-mass radius) for the 2G and the 1G
populations, where $\srad$ and $\stan$ are, respectively, the
projected radial and tangential velocity dispersions as measured on a
plane coinciding with the cluster orbital plane. This is the outermost
region probed by our observational data where the difference in
anisotropy between 1G and 2G populations is strongest. The results of
our simulation show a trend qualitatively consistent with that found
in the observational data: as 2G stars diffuse from the inner regions
(where they formed and are initially segregated), they populate the
outer cluster regions preferentially on radial orbits as shown by the
increasing radial anisotropy of the 2G population. The trend found in
our simulation agrees with what was found in a recent numerical study
by \citet{hb2015}.  The different formation and dynamical history of
1G and 2G stars, which are currently located in the same outer region
of the cluster (in this case between 1.5 and 2 $r_{\rm h}$), is
revealed by the differences in the kinematic properties. These
differences are stronger in the cluster's outer regions, which are
initially dominated by 1G stars.  As the cluster evolves, the
cluster's outer regions are populated also by 2G stars, diffused from
their initial inner location.

The inset of Fig.~\ref{f5} shows the time evolution of the mean of the
ratio of the 1G to the 2G $\srad$ and $\stan$, measured between 1.5
and 2$r{\rm _h}$. The initial differences in the $\srad$ and $\stan$
of the two populations evolve in a direction consistent with the
observed trends:\ the radial velocity dispersion $\srad$ of the two
populations becomes increasingly similar while $\stan$ of the 2G
becomes smaller than that of the 1G and, in agreement with our
observational findings.  It is this difference that is responsible for
the differences in the anisotropy of the two populations.

\section{Conclusions}
\label{sec:conclu}

In this Letter we have presented an \textit{HST} study of the
kinematical properties of the MSPs of NGC~2808 based on high-precision
PM measurements \citep{2014ApJ...797..115B}.  In a recent study,
\citet{2015ApJcc} identified five different populations in this
cluster:\ two of these populations (named D and E) correspond to the
2G populations in the mMS and bMS identified in an earlier study by
\citet{2007ApJ...661L..53P} while the other three populations (A, B,
and C) correspond to the rMS population in the Piotto et al.'s study.
In this Letter we show that the five stellar populations of this
cluster are characterized by differences in their kinematics.

Specifically, we find that in the outermost regions probed (between
1.5 and 2 $r_{\rm h}$) the velocity distribution of the 2G populations
D and E is radially anisotropic (the deviation from isotropy being
significant at the $\sim$3.5-$\sigma$ level).  Our data show that the
larger radial anisotropy in the 2G populations D and E is due to the
fact that these populations have smaller tangential velocity
dispersions than the other populations.  Qualitatively similar trends
have been found by \citealt{2013ApJ...771L..15R} in 47~Tuc. No
significant differences are found in the radial velocity dispersions
of all the populations and between the kinematics of population B and
that of A and C (with only possible hints of a larger $\srad$ of
population A in the innermost region, a larger $\srad$ for population
C in the outermost bin, and a larger $\stan$ for populations A and C
in the outermost regions all at less than 2-$\sigma$ level).

The results of an N-body simulation show that the differences in both
the radial anisotropy and in the tangential velocity dispersion
observed for populations D and E are consistent with being the
kinematical fingerprint of the diffusion of 2G populations from the
innermost regions, where they are initially concentrated, towards the
cluster outer regions. A more extended survey of models, probing
broader ranges of different initial conditions, is necessary to fully
explore the possible evolutionary paths of the 1G and 2G kinematical
properties and their dependence on initial conditions.

On the observational side, we emphasize the importance of extending
the current study to even more external cluster regions. In fact, the
outermost regions are essential for a complete characterization of a
cluster's kinematical properties and, in particular, to shed light on
the possible effects of the external Galactic tidal field.  These
effects are expected to limit the outer development of the radial
anisotropy and make these more isotropic (and possibly even slightly
tangentially anisotropic).

We plan to extend the analysis presented here to other GCs as
possible, using the PM catalogs of \citet{2014ApJ...797..115B} in
combination with the UV photometry of \citet{2015AJ....149...91P}.
This will create a substantial case history that will help us in
better understanding formation and evolution of MSPs in GCs.

\acknowledgments Support for this work comes from STScI grants for HST
programs AR-12845 and GO-13297. E.V. acknowledges support also by
grant NASA-NNX13AF45G. G.P., S.C., F.D'A. and A.R. acknowledge support
from PRIN-INAF 2014 (PI: S. Cassisi).

\end{document}